# Automaton based detection of affected cells in three dimensional biological system

Jitesh Dundas

Scientist, Edencore Technologies (www.edencore.net )
Email addresses:

JBD: jitesh@bioclues.org

## Abstract

### Background

The aim of this research review is to propose the logic and search mechanism for the development of an artificially intelligent automaton (AIA) that can find affected cells in a 3-dimensional biological system. Research on the possible application of such automatons to detect and control cancer cells in the human body are greatly focused MRI and PET scans finds the affected regions at the tissue level even as we can find the affected regions at the cellular level using the framework. The AIA may be designed to ensure optimum utilization as they record and might control the presence of affected cells in a human body. The proposed models and techniques can be generalized and used in any application where cells are injured or affected by some disease or accident. The best method to import AIA into the body without surgery or injection is to insert small pill like automata, carrying material viz drugs or leukocytes that is needed to correct the infection. In this process, the AIA can be compared to nano pills to deliver or support therapy. These small automatons nevertheless called as small pill-sized robots could then be inserted into the body via the mouth and then made to search for all affected areas. We propose that affected areas can be found by means of an algorithm that uses artificial intelligence-based spiral search techniques in vertical and horizontal directions.

**Results**

We believe the automatons may be tracked and controlled externally using sensors, lasers or sonography, thereby controlling sensors present in them. Furthermore, this may be used to transmit information while probabilistic measures of location and the extent of infection could be rendered to each cell in a given area. NanoHive simulation software was used to validate the framework of this paper. The existing nanomedicine models such as obstacle avoidance algorithm based models (Hla K H S et al 2008) and the framework in this model were tested in different simulation based experiments. The existing models such as obstacle avoidance based models failed in complex environmental conditions (such as changing environmental conditions, presence of semi-solid particles, etc) while the model in this paper executed its framework successfully.

**Conclusions**

Come systems biology, this field of automatons deserves a bigger leap of understanding especially when pharmacogenomics is at its peak. The results also indicate the importance of artificial intelligence and other computational capabilities in the proposed model for the successful detection of affected cells.

# Background

**Introduction**

### 1.1 Cell is the basic unit of life

A cell is the basic unit of life. It is a complex system working independently as well as in symbiosis with its external environment (Bolsover et al 2003). In humans, the cells are differentiated based on the functions they carry out. For example, the red blood cells (RBC) or erythrocytes, leukocytes, lymphocytes etc. It is believed that several cells metamorphose to different functions for example, transmission of electrical signals is done by neurons, transport of oxygen by red blood cells, destruction of infecting bacteria through macrophages, contraction by muscle cells, chemical processing by liver cells etc.

### 1.2   Intelligent tools are required to combat cellular infection

More background is needed why automatons are a must and what role they may have to play.

### 1.3 What are Automatons?

A nano-sized artificially intelligent robot that can investigate biological environment is termed as automaton. Automatons are singletons and .are made of??? . On the other hand, the micro-motor approaches s automaton in structure and functionality (Friend 2009) which therefore can be made artificially intelligent to execute the course of action. First, cellular infections generally spread in clusters or in roughly circular regions.

**Applications of automatons:**

1)     The automaton is used to move into the body and detect any affected cells.

2)     This automaton can be very useful in drug-delivery.

**Objectives of our study:**

1.   Dead human bodies or small animals (resembling humans in their body structure and functions) will be used as subjects rather than live humans.

2.   The framework will be first implemented on a part of the body e.g. the gastro-intestinal track. Again, we will try to employ multiple automatons here to reduce the time lag in recording the information. Also, it is possible that the entire automaton may not be able to detect all the cells in the region as each region is very dynamic and difficult to track. The gastro-intestinal track is a difficult terrain to walk through the gastro-intestinal track lumen and thus all cells may not be able to scan all the cells. IN such cases, it is advisable to divide each region for separate automaton for scanning the region. Thus, for each different type of terrain of the region, a different automaton will be used.

3. The automaton will move along a linear trajectory. However, the search mechanism used by the automaton will follow the algorithm described below. In this paper, any reference to search mechanism will mean the search mechanism of the automaton and not its movement.

4. The automaton will have high resolution cameras that will be able to scan in all directions, especially in a circular fashion, while it moves forward recording the cellular details of the region. In short, while the movement of the automaton will be linear, the movement of the cameras of the automatons or any of its parts used in recording cellular details will be circular, in horizontal and vertical directions. This is because body parts generally have closed cylindrical structures.

5. Please note that detection and analysis of each cell will take time and the scan of the entire region will be very time-consuming. Thus, it can be useful to employ multiple automatons to scan each region independently.

6. It is possible that one type of infection might share similar traits with another infection. Thus, it will be very important for the automaton to have knowledge of 2-3 unique features of each infection. This will help it to record information for each infection clearly.

7. The automaton will be embedded with artificial intelligence and logic so that it can perform its expected work without any human interference.

Freitas (Freitas R. 1999, Nanomedicine) pioneered the concept of nanomedicine, when he published the first design on nanorobot. He has extensively covered several aspects in Nanomedicine in his work, including inter-nanorobot communication, locomotion and applications. He has thrown light on biocompatibility issues in Nanomedicine too. The use of an artificially intelligent automaton is a step towards extending the work on Nanorobotics that has been done by Freitas, Cavalcanti and other eminent experts in this field. Cyril Ng (Cyril Ng et al 2008) has suggested the use of nanorobots in the tracking/treatment of targeted cells in the body. There is a need for models (Gutierrez 2009) that can help us find and analyze all the affected cells in the human body. Such models are very important for the efficient use of AIA. According to Sitti (Sitti .M. 2009), the future challenges in developing micro-robots include finding methods for interacting with them inside the body. Zhang (Zhang et al 2009) have reported artificial bacterial flagella that simulate their natural counterparts and swim in blood. Cavalcanti (Cavalcanti A. et al ICARCV 2006) have proposed the use of a nano-robot that uses a "follow gradient with attractant signal" method to detect cancer cells. Witten (Witten 1989) has discussed the growth modifying factor (GMF) in detail in order to show its impact on cancer growth. Codourey (Codourey et al 1994) had proposed the design of nanorobots with focus on high accuracy and reduction of overall dimensions. Cyril (Cyril Ng et al 2008) has emphasized on the use of Evolutionary Nanotechnology in biological applications using artificially intelligent materials. Cavalcanti (Cavalcanti A., Shirinzadeh B., et al AWFM06 2006) has proposed a framework for nanorobots hardware in laparoscopic cancer surgery. He also mentions that Preoperative lymph node staging with computerized tomography or magnetic resonance imaging (MRI) was not successful as sensitivity and specificity are limited. Zesch (Zesch .Z., Buchi R., et al 1995) and colleagues have discussed two piezoelectric devices for positioning in the nanorobot. Mallouk (Mallouk T. et al) presents a nano engine that was two-micron-long gold-platinum rods which could move ahead in a solution of water and hydrogen peroxide (H2O2) by pushing the fluid along their sides. Greitmann (Greitmann G. et al 1996) proposes a micro machined gripper system with integrated sensor and actuator. Cavalcanti (Cavalcanti A. 2005) has emphasized the need of automation of atomic-scale manipulation in nanotechnology applications. Ferreira (Ferreira A. et al 2005) and colleagues have suggested the use of simulated nano-environments in virtual technologies. Requicha (Requicha A. 2002) initially discusses the construction of nanorobots and then later focuses on nanoassembly by manipulation with scanning probe microscopes (SPMs). Duncan (Duncan R. 2004) discusses the possibility of nanotechnology applications in the treatment of diseases. The author too proposes the concept of nanorobots being used for drug-delivery and that this could be practically implemented in future. Casal (Casal A. et al 2003) discusses the concept of a nanorobot in a simulated microenvironment that can move in semi fluid areas. This nanorobot simulator uses a set of design parameters. Cavalcanti (Cavalcanti A. et al 2006) have proposed an excellent computational method for developing nanorobots in liquid environments, using Reynolds number, for medical applications like drug-delivery and surgery. Sitti (Sitti M. 2009) has talked about nanobots or tiny robots that can be injected into the body to perform medical procedures. Setti also cites the importance of methods for the nanorobots to be used effectively in biological systems for medical applications. Hamdi (Hamdi et al 2006) proposes a simulation for bio-nanorobotic prototyping using NAMD software and VR techniques. The authors propose to use these biomolecular motors for manipulation and structuring at the protein level. In an innovative paper, Lewis (Lewis et al 1992) has proposed a method for coordinated control of a large number of nanorobots. Sudo (Sudo et al 2006) have proposed a magnetic swimming mechanism in nanorobot development. The movement of proposed robot is via the oscillation of its tail. The authors also mention that the miniaturization of the permanent magnet is needed for the robot to move through the smaller capillaries. In their reply to the doubts raised by Curtis (Curtis A. 2005) on their paper, Cavalcanti (Cavalcanti A. and Freitas R. 2005) have mentioned the development of digital circuits in living cells. In an excellent paper, Bedau (Bedau et al 1997) propose a quantitative comparison of such trends in model systems and in the biosphere. Cavalcanti (Cavalcanti A. and Freitas R. Aug 2002) have proposed the design and simulation of a team of multipurpose nanorobots for activities like nanomedicine delivery at affected cell regions. Brunner (Brunner M. and Stemmer A. 1998) has suggested the design and control of a nanorobot with 2 - dimensional linear motor and AFM (atomic force microscope) for image processing of affected regions. In another excellent paper, Alouges (Alouges F. et al 2007) has proposed a numerical algorithm to compute optimal strokes for swimming of artificial micro-swimmers. However, they also emphasize the need for more mathematically abstract tools for biological swimmers. Ion (Ion R. and Cocina G. 2008) in their paper studied the controlled aggregation of meso-5, 10, 15, 20-sulfonato-phenyl porphyrin (TPPS4) at room temperature. using electron microscopy (SEM), transmission electron microscopy (TEM) and UV-visible spectroscopy. The authors have used nanorobots for application in brain aneurysm. Saniotis (Saniotis 2008) discusses the future applications of nanotechnology. He throws light on nanocosms, Respirocytes and microbovores as possible future examples of use of nanotechnology. Freitas (Freitas .R.2003) has pioneered the concept of respirocytes and microbovores in the field of nanomedicine. Hla (Hla et al Nov 2008) in his paper has proposed the particle swarm optimization algorithm (PSO) besides obstacle avoidance to control nanorobots movement in the human body. The authors propose a 2-dimensional coordinate system to implement the obstacle avoidance algorithm. Pitt (Pitt J. 2008) discusses the possibility of self-replicating nanorobots that could cause unwanted harm for the human race. Cavalcanti (Cavalcanti 2003) has proposed a noel approach in nanorobot control design for assembly manipulation using graphical simulations. Again with this co-authors (Cavalcanti et al CIMCA 2006), he proposes the use CMOS based chips for the manufacturing nanorobots.

## Methodology

### 2.1 Framework explanation

First, there is a need to ensure the better functioning of AIA, which is the primary motivation for this paper. It is also necessary to create a support mechanism that will enable AIA to be used for delivering more effective treatment to patients.

We consider cells as points, represented in Cartesian coordinates by P(x, y, z) where x is the length, y is the breadth and z is the height of the cell location in a three-dimensional human body map. The automaton then assigns a probability, called the Cellular Infection Probability Measurement (CIPM), that the cell being affected. Cellular representation and comparisons require mathematical techniques to identify the affected regions in the body. As the cells are analyzed by the automaton, the resulting information can be plotted in a graph. This will be a two-dimensional graph called the Cellular Infection Probability Graph (CIPG). This graph can be plotted with its X axis showing the cell location P(x, y, z) and the corresponding Y axis showing the probability (CIPM) of that cell being affected.

For each cell point under investigation, there is a measurement assigned called Cellular Infection Level Measurement (L). This level gives a value of the degree or level of infection present in the cell. This degree of infection is found by finding the average of the values for attributes of the infection determinants (on a scale of 1-10) like color, cell width, cell height, cell length, presence of unwanted particles or pathogens, etc. As any cell has more than one trait (like cell color, enzyme content 'XYZ'), we denote it by $L_{n, m}$, where m is the index of the trait for the cell n. Each trait is measured on a scale of 1-10 and then the average of all the trait measurements is taken to get the Average L (Avg ($L_n$)) value. Here, the measurements of each trait $L_{n, m}$ and Avg ($L_n$) is taken. For simplicity, we will refer Avg ($L_n$) as $L_n$.

Please note the difference between the two measurements, CIPM and L. CIPM is the probability of finding a cell as being affected. For e.g.) when the automaton finds a cell point, it has to decide its type i.e. healthy type of cell or affected type of cell. The automaton will have a sample of the affected cell and based on this sample (or knowledge of this cell that could be stored in its memory), it can make a comparison. It will assign a probability that this cell has a 0.4 or 0.8 probability of being affected. This is similar to the case when a doctor tells the patient that the patient has 40% or 80% probability of survival after the operation. This measure has been kept to let us know which cell is probably affected and which is not.

L is a different measurement. It measures each of the actual traits of the cell like cell colour, protein content, etc. These traits need to be measured too in order to give us more information about the cell. This information allows better choice of treatment. This paper should help the development of better implementation and optimum utilization of AIA in the body. To achieve this goal, mathematical and computational techniques have been employed.

It is evident that AIA need to be incorporated into the body orally or via injections. However, to ensure that these automata reach the affected regions in the best way possible, we need to deploy an automaton (next-generation pill with its own logic i.e. a robot), that can carry and control the entire process of material transfer and damage repair. For e.g.) .Also, we need to ensure that the patient undergoes minimum pain and stress due to the incorporation of AIA. Using recent advances in robotics and micro-technology, we can easily embed the necessary logic into the transporter, which will be of the size of a pill. The automaton will be controlled externally via using sensors or chips present in it. We can control and command the automaton and see how it is working inside the body. We could call such automata next-generation pills, as they can control and improve the logic and manner using the requisite material.

The entire framework can be represented in the following steps (Figure - 6):-

1.  Travel in the 3-dimensional biological system and study each cell one by one.

2.  For each cell n, measure the probability CIPM (or Pn) that the cell is affected.

3.  Based on the value of P (n), we get the value of $L_{n,\,m}$. Pass on the values to the database. This database is outside the biological system for our records.

4.  In the graphs CIPG and CIWG being calculated, plot the cell's P (n) and $L_{n,\,m}$ for each type of affected cell found.

5.  Pass the treatment material from the automaton to the affected cell or trigger the repair mechanisms for the same.

6.      Send the feedback or any other details to us (external environment) for our records. The logic and direction for the automaton can be updated or changed as per our needs. This includes any exception handling.

7.      Repeat the steps from Step-1 to Step-6 for each cell. Send any information needed to be given to the automaton for recording further cells.

Techniques like Laparoscopic and Open Colorectal Surgery are possible methods for use in this framework. Braga and colleagues (Braga et al 2002) have experimentally shown that laparoscopic surgery is better than open surgery. However, capsule endoscopy is proposed here as it is involves the use of camera like pills for studying the affected regions. Capsule endoscopy is a technique in which a small pill-like camera is injected into the human body. This pill-like camera then takes pictures of the internal regions of the body. This technique is widely used in obtaining images of the gastro-intestinal track. The automatons can be fitted with such high-resolution cameras for scanning each cell.

The modelling of the biological system (human body) is important as it gives us an idea about the expected location and possible type of cell at each point. Thus, we can safely deduce that if there is a possible dislocation or deviation of the expected cell found by the automaton at that point, then it is possible to find an affected cell/external agent at that point.

## 2.2   NanoHive and simulation based experiments

As shown in figure 7, NanoHive simulation software was used to validate the framework of this paper. IBSEAD (Dundas .J. and Chik D. 2010) and adaptive neural networks were used as machine learning algorithms to help provide the intelligence in this framework. The existing nanomedicine models (such as obstacle avoidance based model) and the framework in this model were tested in different simulation based experiments. Obstacles were introduced in varying capacities to test how the automaton would perform in the framework.

The use of IBSEAD is justified as the presence of complex environments requires the handling of unknown entities, which is not done by other machine learning algorithms. The experiments were carried out using proven simulation software based scenarios, after an initial round of dry-run algorithmic analysis of the scenarios.

## Discussion

### Search Mechanism for Automaton to find affected cells in the body

Any cell can be distinguished from all others on the basis of its location and functionality. In order to find the affected cells, we need to treat the human body as a 3-dimensional biological system. The human body (Gray 2005, B J C Perera 2004) is symmetrical but complex in structure, and needs a three-dimensional approach to structural analysis. Every point or cell in the body is identified in terms of its x, y and z coordinates. These correspond to the distance of the cell (from the origin O) in the x, y and z directions. Thus cell P is given by P(x, y, z). Now, the search mechanism for affected cells cannot be executed linearly, since those affected cells do not spread and are not organized linearly. They are randomly distributed and thus need to be searched using an algorithm that takes account of such randomness (Figure-3). The internal regions are mostly cylindrical in structure like a hollow pipe. Thus, the cells have to be tracked initially in a vertical (standing) spiral-like movement, like a roller coaster. However, we also need to know whether an affected cell is isolated or part of a bigger cluster of affected cells. A probability of infection can be assigned to each cell, the Cellular Impact Probability Measure (CIPM), by comparing the sample of affected cells. We also need to know the degree or extent of infection in the cell. This can be obtained by studying attributes (colour, size, thickness, materials present, etc.) and finding the extent to which each differs from the healthy cell. Each $L_{n, m}$ is measured (scale 1-10) and then an average of all attributes is obtained. This measurement will be called Cellular Infection Level Measurement (L) for any cell. It is also denoted by $L_n$ as shown in the previous sections. The CIPM and L tell us if a cell is affected and define the type or extent of infection. Any region may have two or more types of infection, so these measurements will give us information about the number of affected cells and the type of infections present in a region. The grading systems (Schulz 2005) for measuring the malignancy of cancer like 'G Grading ', can be customized for serving the same purpose in this model. The L can include the values like those from 'G Grading' system for measuring the malignancy of the affected cells.

As diseases such as cancer generally occur in clusters, it would be better to track affected cells in clusters. Thus, when an affected cell is found, the search method will change. It will now become a spiral-like search with horizontal movement. The starting point will be the currently tracked affected cell; taking this as the centre, the search will proceed spirally in a horizontal direction. This will help to find clusters of affected cells, if any, in that region. The following steps (Figure-3) outline the method by which the automaton will locate and record the affected cells in the human body.

1. Spiral search in vertical direction:-

2. Spiral search in horizontal direction:-

These steps have been explained with an example in the supplementary material, showing both the types of search algorithms working in sync to detect the affected cells. The spiral search in vertical direction will in execution till the first affected cell is detected by it ($Pn > 0.5$). Once this happens, the spiral search in horizontal direction will start executing, with the process executing till the cells that are detected become non-affected i.e. $Pn < 0.5$. Once this happens, the spiral search in vertical direction is started again from the point where it had left. Please note that this includes ignoring the cells that have already been scanned and recorded. Further details are present in the example mentioned in the supplementary material of this paper.

Once all the cells have been studied and their probabilities measured, we can create a graph (**CIPG**) to reflect the spread and impact of the affected cells and their underlying cause. A two-dimensional graph (Figure.-2) can be plotted with the x coordinate showing the cell location P and the corresponding y coordinate showing the probability measure (**CIPM or Pn**) of that cell being affected or injured. The graph will portray the number of affected regions and the extent to which cells are affected throughout the body. This is very useful in finding the location and type of cancer cells in the body. As cancer is difficult to predict and track, this model can be very useful in the control and treatment of the same.

Also, for each point under investigation, there is a level assigned called cellular infection level(L). This level gives a value of the degree of infection present in the cell. This degree of infection is found by finding the average of the values for attributes of the infection determinants (on a scale of 1-10) like colour, cell width, cell height, cell length, presence of unwanted particles or pathogens, etc. We also need to measure each of the traits measurement like $L_{n,m}$ where n is the cell and m is the each of the cell trait. This measurement $L_{n,m}$ is taken for each cell trait. Later the average of each of these traits is taken to achieve Avg $L_n$ (or L).  For e.g. if there is a cell with n=12340(index of the cell) and coordinates P(100,200,300 ) ,  then we represent the cell by $W_{12340}$. Assuming that the cell 'n' is having CIPM (or P(12340) = 0.7. Let us consider the traits like cell colour and enzyme 'XYZ' (assumed name) content. Thus 'm' will have index values 0 to 1. Thus, we have cell colour trait as $W_{100,0} = 0.7$ and enzyme 'XYZ' content as $W_{100,1} = 0.6$. We can extend 'm' as per the number of traits considered for the cells. Thus, the average of cell n=12340 is given by:-

$$W_{12340} = (0.7 + 0.6) / 2 = 1.3 / 2 = 0.65.$$

This will help us to propose better, methods and techniques of treatment to control the affected cells for the underlying cause. Doctors and other experts will obtain a better picture of the disease or infection, enabling to prescribe better medications and management procedures for the patient.

There is a point to be noted in this search algorithm. It is possible that a cell that was recorded in the horizontal direction based spiral search is again found in its path. In such a condition, such a cell is just skipped by the automaton and the search proceeds to the next cell. For e.g.) suppose that a cell (10, 20,100) was encountered by the automaton in the horizontal search and it is recorded as affected with a cellular infection probability of 0.6. When the vertical direction based spiral search is encountered again by the automaton in its path, then this cell is skipped by the automaton. Thus, the search proceeds ahead with the next cell.

It is also possible that there are two or more types of infections or external objects (particles, pathogens, etc) present in the area of investigation. Thus, there is a chance that the automaton may not be able to decide if such an object or affected cell is actually affected or healthy. The automaton can decide based on the comparison between the affected cell and the sample of the healthy cell that was in the beginning of this framework. Thus, in such a case, the points in the investigation area are assigned a probability of zero and the cellular infection level as per the values obtained by the automaton. The CIWG graph will show the actual values based on the measurement obtained by the automaton. Thus in the CIPG graph, such a cell is present on the X-axis (Figure – 5). Also, any new affected cell or external object must be recorded on a separate CIPG graph, with the curve beginning from the point where it was found. Thus, when such a new infection is recorded, the automaton will compare the current cell against the two sample cells (or objects). It will first compare the current cell against the original sample of the cell and record the impact and other details on the CIPG graph. It will again do the same for the new type of affected cell or the external object. Thus, two curves will be plotted against the cell points in the CIPG. Also, for every cell that is found to be of the second type of infection, we can assign a cellular infection probability of zero on the first curve. This value of zero will tell us that the cell at this point is of the second type of infection.

This adjustment of creating a new curve on the CIPG is essential in understanding the possible relationship between the two types of infections (or objects) present in the investigation area of the body. For e.g.) the first type of infection maybe an infection in the urinary tract, while the second type of infection cell may be a small calcite stone in the kidney. Thus, plotting two CIPG graphs will give us an idea of:-

1. How many affected cells of each type are present in the area of investigation?
2. What is the possible relationship between the types of affected cells?

It is also possible that the cell may have both types of infection present on it. For e.g.) a cell showing a pus cell may have a kidney stone particle engulfed in it. Thus, both these cells will be assigned the same cellular infection probability. Such a case clearly implies that there is relationship between the two types of infection found in the cell. For e.g.) a kidney stone particle can cause an infection in the urinary tract. Also, continuous problems in digestion or lung-related issues may point to stomach or lung cancer respectively.

This adjustment in the search mechanism can be generalized to include more than two types of infections in the investigation area. The method of recording the different types of infection will remain the same as shown in the above paragraph. Such an adjustment can help us find the root cause of infection or medical condition in the patient. This will in turn help us find the most suitable treatment of the patient.

In this paper, the information that is collected ($P_n$ and $L_{n, m}$) by the automaton is stored in a database. This information can be analyzed t obtain critical insight into the condition of the cell.

It is also possible that the automaton may face an external agent or obstacle in its pathway. In such a case, the automaton will treat this agent or particle as a new type of affected cell and thus it will create a separate curve on the CILG graph for the same. Also, it may be possible that the automaton may face an issue or obstacle in its path. In such cases, it can communicate with us via the sensors for directions.

**Measuring the traits of the cells**

Each cell is made up of several traits in the body. For finding if the cell is affected or not, we need to measure each cell to know the level to the cell has been affected. In order to calculate the value L for each cell. We need to have generalized formulae to do so.

Lets us consider a cell $L_{1,m}, L_{2,m}, \ldots\ldots L_{n,m..}$ If we consider that each cell has 10 traits that could cause or indicate cancer. Some of these include oncogenes content, cell colour, etc. Thus, each of them would be having their own scale of measurements such as kg, mtr, etc. Thus, these unites have to converted into relative 1-10 scale to make them in sync with the framework and thus comparable with other cell traits. Once all the traits are measured, their average can be found to obtain $L_{n.}$ This average is shown along with the individual traits as curves on the CILG graph to get a clear picture of the cell conditions in the body.

It is a very important to note that the traits that are to be measured may be difficult to convert into relative 1-10 scale. These traits may be in decimals to retain the accuracy in calculation. Thus, the conversion procedure needs to be tested and verified before being used with this framework.

**Design of Automaton**

It is proposed that the automaton be designed on the lines of the structure of the sperm cell. The sperm cell (Gray 2005) is one of the most fantastic cells that can travel a long distance on its own towards its target. Next, it also pierces into the egg cell and releases its chromosomes into the egg. This is a very interesting model used by the sperm cell and the automaton of this framework is inspired from the same.

The automaton will have a sperm-like structure with the head containing the camera, radar and processing unit. The camera will be used for scanning and detecting affected cells. The radar will be used for tracking and capturing the signals from other automatons if needed. It is also proposed that the automaton should be made capable of capturing any errors in cellular signalling. With the future advancement of nanotechnology, this automaton will hopefully be able to provide correction signals for removing any errors in cellular signalling. The processing unit will be responsible for handling the decision-making and instruction execution for the automaton. The middle part of the automaton will consists of the engine for navigational propagation, communication equipment and medicinal material to deliver to the affected cell. The communication equipment will consist of processing the incoming signals from the radar and sending the signals to the other cells via the radar as well. One point of debate still revolves around the choice of electrodes for creating fuel for the automaton. In some previous scholarly papers, gold electrodes have been proposed. However, the choice still remains with the scientist executing this framework. The tail of the automaton will act as a propellant for propagation. The automaton will have to be highly connected so that communication between its counterparts as well as external cells and pathogens can happen without any problem.

The body of the automaton should be biodegradable so that even on self-destruction, it will not harm the body.  Once the automaton will reach the affected cell, it will have to release the medicine for treatment. Again, after its task is completed, the automaton will be released from the body along with the body waste material. The automaton must degrade into powdered granules so that the particles can be released from the body without any problems for the host body. The locomotion of the automaton, driven by the tail and the directions from the head of the automaton, will help in the free movement of the latter. Again, the automaton will move in the manner of a snake-like movement with a zigzag-body movement based form of locomotion. This will resemble a snake moving on sand and the speed will be highest if the head of the automaton is smooth enough to cut through the liquid of the hosy body. Please note that the automaton's head has to be very strong as it will encounter the maximum hurdles and direct accidents with the obstacles in the automaton's path.

There may be a need to deploy multiple automatons with each automaton working in a specific region of the body. Thus, multiple networks based architecture, with each network having multiple automatons, is proposed for this framework. Here, each network of automatons will work in a specific region. There will be a controlling automaton that will control the other automatons in the network.

The automaton will have the tip part of the former filled with poison or necessary weapons in order to fight the pathogens or other unwanted obstacles that cannot be avoided, negotiated or compromised.

It is not necessary that the framework must follow the design of the automaton mentioned in this paper. The framework can be deployed for any automaton design that could be available at the time of execution of the same. Also, the design of the automaton can be further enhanced using the existing advances in nanotechnology.

The automaton (Figure-6) maybe given the ability to self-detect and evolve to become smart enough to analyze the host body environment. The automaton will need artificial intelligence and other search methods to record all the stored data. The automaton will also benefit from the processing of information made available from the analysis done previously so as to help make better decisions in its execution of current responsibilities.

**Advantages and Support for Argument:-**

1.    The mechanism by which the automaton searches for affected cells cannot be executed linearly, since the affected cells are not spreading. They are randomly distributed so they need to be searched using an algorithm that takes account of such randomness.

2.    This method takes into account the presence of clusters of affected cells and also those cells those are minute and isolated. Both these extreme cases will be addressed.

3. This paper seeks to identify the best way to employ the AIA or other types of cells, in resolving and preventing the problems of cancer cells and their spread.

4. Other methods employed for searching for affected cells do not use such a combination of linear and spiral techniques. They tend to focus more on linear and tree-based A.I. techniques.

5. This paper outlines those aspects of support that tend to be overlooked or minimized during Artificial Intelligence research. Both the mechanism and the usage procedure should be considered equally if the problem and its cure are complex.

6. This paper also allows us to record and analyze the root cause of the infection present in the body.

7. This will help us in better treatment for the patient.

## Results

The results indicated that the use of machine learning concepts helped in improving the performance of the automaton by in complex environments and in the presence of dynamic obstacles. The existing models such as obstacle avoidance based models failed in complex environmental conditions (such as changing environmental conditions, presence of semi-solid particles, etc). Moreover, the information collection and self-adaptation helped this framework helped the automaton to change its behaviour, especially in the presence of obstacles. When the obstacles were found, the automaton (using IBSEAD and adaptive neural networks as machine learning algorithms) learned about the obstacles while avoiding them at a comfortable distance. When another similar obstacle was found, the automaton used the past experience and knowledge to handle the situation. This predefined ability to handle situations and obstacles were an important aspect in the success of the framework. As shown in figure 8, the results indicate the higher success ratio of the paper's framework when compared with other learning algorithms. Moreover, the used of analysis and graphs such as CIPM helped in better coverage and detection of multivariate affected cells. Thus, the framework in this paper is found to be much better in performance than the existing models.

This theoretical paper is intended to enhance the function of AIA by defining the carrier framework and the search for identifying all the affected cells in the body. There are applications to cancer as well as to traumatic accidents to the body. In future, this framework can be extended to study and analyze the internal working of any cell. Also, it can be extended to study of the adaptation techniques in the internal parts of cells using automata. This framework may be utilized productively to analyze mutations and find mechanisms for repairing them. MRI and PET scans can find the affected regions at the tissue level.

However, we can find the affected regions at the cellular level using this framework. It would be interesting to study the framework under a 4-dimensional biological system. The fourth dimension in this system will be time (in seconds). Thus, we can get the analysis of the cells in the human body at different points in time.

There are several enhancements in this framework that can be implemented to make its utilization eve better.

This design can be further enhanced by adding repair and medical-support system logic into it. Thus, based on information from the graph, decision support systems for doctors and biologists (to suggest prospective repair and solution strategies and mechanisms) can be integrated to aid in the treatment of the patient.

The automata will perform several activities, of which the main ones are (Figure-1). This framework has many possible applications. It can also be used to remove thrombi. It can help us obtain clear images using Magnetic Resonance Imaging (MRI) and Positron Emitting Tomography (PET). In addition, the automaton can be modified with devices to remove affected cells or carry materials for treatment. One example would be the detection and removal of cancer, which is generally difficult with current medical treatment.

The application of oncogenes (Crocker et al 2003) in the treatment of some cancers can be aided with the help of this model. Changes in the oncogenes can lead to cancer. This can happen through a number of mechanisms that are not unique to any one gene. These mechanisms can be due to alteration in the protein itself, over-expression of the protein, or loss of control mechanisms. We can study these mechanisms and analyze how they can be stopped or reversed, in order to treat cancer.

Other applications include the removal or treatment of following diseases:

1.    Dissolving clots caused in the "cholecystitis".

2.    Removal of small abnormal tissue segments in uterine fibroid disease.

3.    Removal of thrombi resulting from deep vein thrombosis.

4.    Treatment of "o**esophageal varices**" by applying the oxidant agent, used for treatment, via the automaton.

This paper is intended to enhance the function of AIA by defining the carrier framework and the search for identifying all the affected cells in the body. There are applications to cancer as well as to traumatic accidents to the body. In future, this framework can be extended to study and analyze the internal working of any cell. Also, it can be extended to study of the adaptation techniques in the internal parts of cells using automata. This framework may be utilized productively to analyze mutations and find mechanisms for repairing them. MRI and PET scans can find the affected regions at the tissue level.

However, we can find the affected regions at the cellular level using this framework. It would be interesting to study the framework under a 4-dimensional biological system. The fourth dimension in this system will be time (in seconds). Thus, we can get the analysis of the cells in the human body at different points in time.

## Conclusions

This paper is intended to enhance the function of AIA by defining the carrier framework and the search for identifying all the affected cells in the body. There are applications to cancer as well as to traumatic accidents to the body. In future, this framework can be extended to study and analyze the internal working of any cell. Also, it can be extended to study of the adaptation techniques in the internal parts of cells using automata. This framework may be utilized productively to analyze mutations and find mechanisms for repairing them. MRI and PET scans can find the affected regions at the tissue level.

However, we can find the affected regions at the cellular level using this framework. It would be interesting to study the framework under a 4-dimensional biological system. The fourth dimension in this system will be time (in seconds). Thus, we can get the analysis of the cells in the human body at different points in time.

## Terms and Definitions

1.  n: - It is the cell represented by the coordinates x, y and z.

2.  O: - It is the point of origin in the 3-dimensional biological system.

3.  x: - It the distance of the cell 'n' in the X-axis of the biological system.

4.  y: - It the distance of the cell 'n' in the Y-axis of the biological system.

5.  z: - It the distance of the cell 'n' in the Z-axis of the biological system.

6.  m: - It is the index of the trait of the cell. For e.g. there are several traits of cell like cell colour, enzyme content, etc. Each of the 'n' cell traits is indexed by 'm'.

7.  CIPM: - Cellular Infection Probability Measurement. The automaton then assigns a probability, called the Cellular Infection Probability Measurement (CIPM), that the cell being affected. It is also denoted by Pn.

8.  L: - Cellular Impact Level Measurement. For each cell point having more than one trait under investigation, there is a measurement assigned called Cellular Impact Level Measurement. This measurement is 2-dimensional for each trait as it is indexed by the cell position 'n' and the cell trait 'm' i.e. it is measured for each cell trait. This measurement gives a value of the level upto which the cell is affected. As any cell has more than one trait (like cell colour, enzyme content 'XYZ'), we denote it by $L_{n, m}$, where m is the index of the trait for the cell n. Each trait is measured on a scale of 1-10 and then the average of all the trait measurements is taken to get the Average L (Avg $(L_n)$) value. Here, the measurements of each trait $L_{n, m}$ and Avg $(L_n)$ is taken. For simplicity, we will refer Avg $(L_n)$ as $L_n$.

9.  CIPG: - Cellular Impact Probability Graph. The graph shows the values of CIPM (or Pn) for each cell 'n'.

10. CILG: - Cellular Infection Level Graph. The graph shows the values of L (or $L_{n, m}$) for each cell 'n'. In this graph, the measurements of each trait $L_{n, m}$ as well as $L_n$ is represented on a separate curve.

## Competing interests

No conflicting financial interests exist

## Authors' contributions

JBD was the sole investigator and author of this manuscript.

## Acknowledgements

The author, J.B. Dundas would like to thank his family and friends without whose support; this paper could not have been completed. The author would also like to thank Mr. Joshua K Hicks and Dr. Yogesh Narkhade for his help in my research. A special mention of gratitude to Prof. Uma Srinivasan and Prof. Bob Bruner. The author would also like to thank Mr. Robert Freitas for his expert comments on this paper.

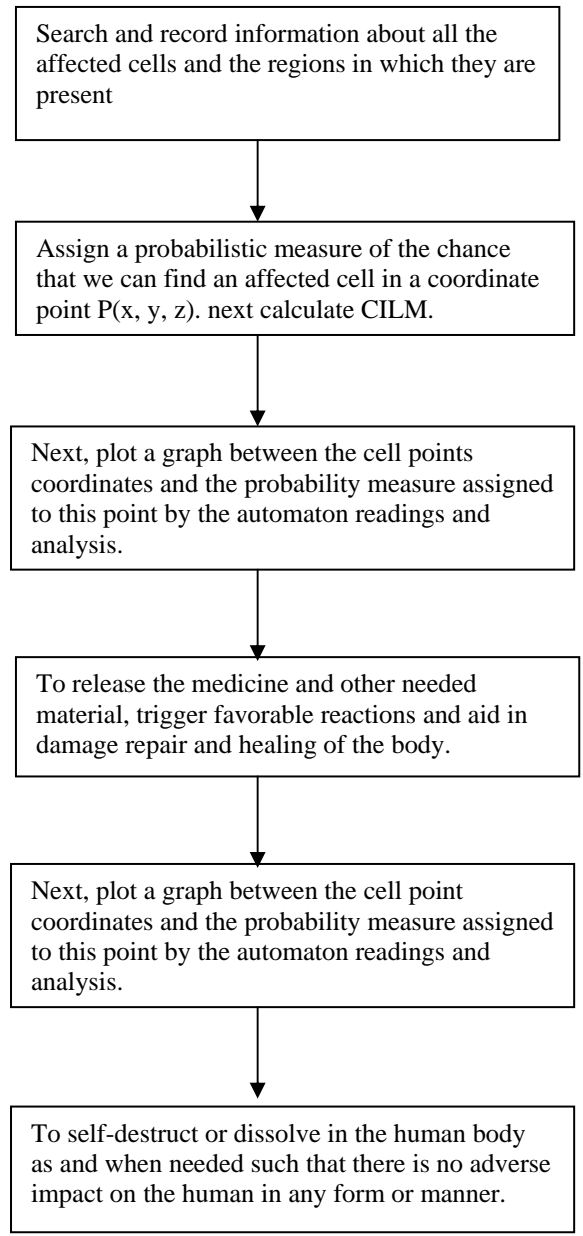

Figure 1) Functions, activities and application of the ADIBS framework using automata

Figure-2) The cellular impact probability graph (CIPG)

Probability (CIPM) is shown on the y-axis and cell point P on the x axis.
This graph area shows the cells in the human body that are affected or injured. Also, note that the point where the infection of another type exists, a corresponding probability of zero is plotted on this CIPM graph for the corresponding cell point. For e.g. in this CIPG graph the infected cell at 1000 is of probability zero as it is of different type of infection. The same cell point will have a probability of 0.5 in the other infection graph as shown in Figure-5). However, the CILM graph will remain the same for all the CIPM graphs.

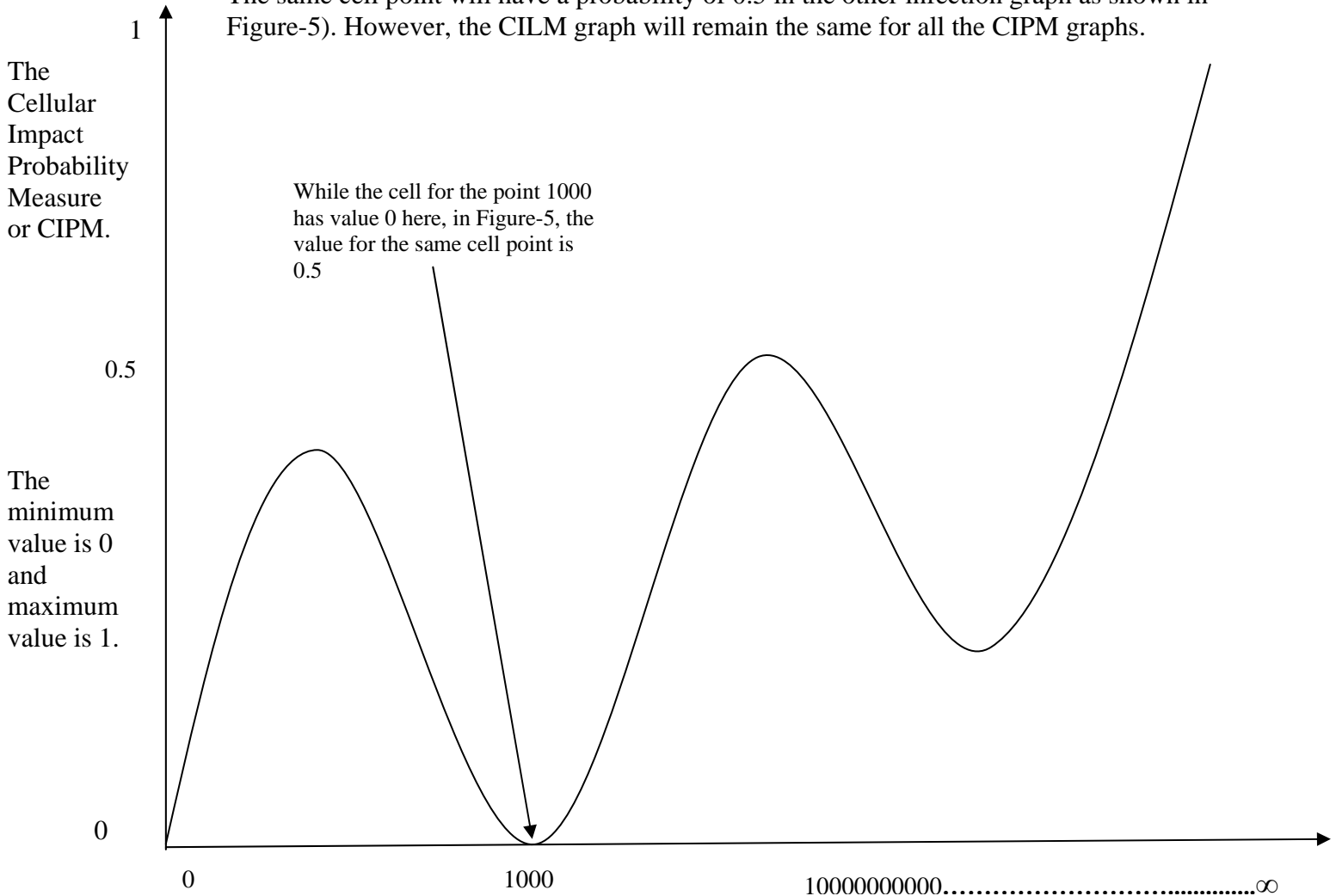

The Cellular Impact Probability Measure or CIPM.

The minimum value is 0 and maximum value is 1.

While the cell for the point 1000 has value 0 here, in Figure-5, the value for the same cell point is 0.5

0.5

0

0        1000        10000000000...................................................∞

Each point P represents an individual cell in the body.

# Figure 3) Search mechanism for finding and recording infected cells

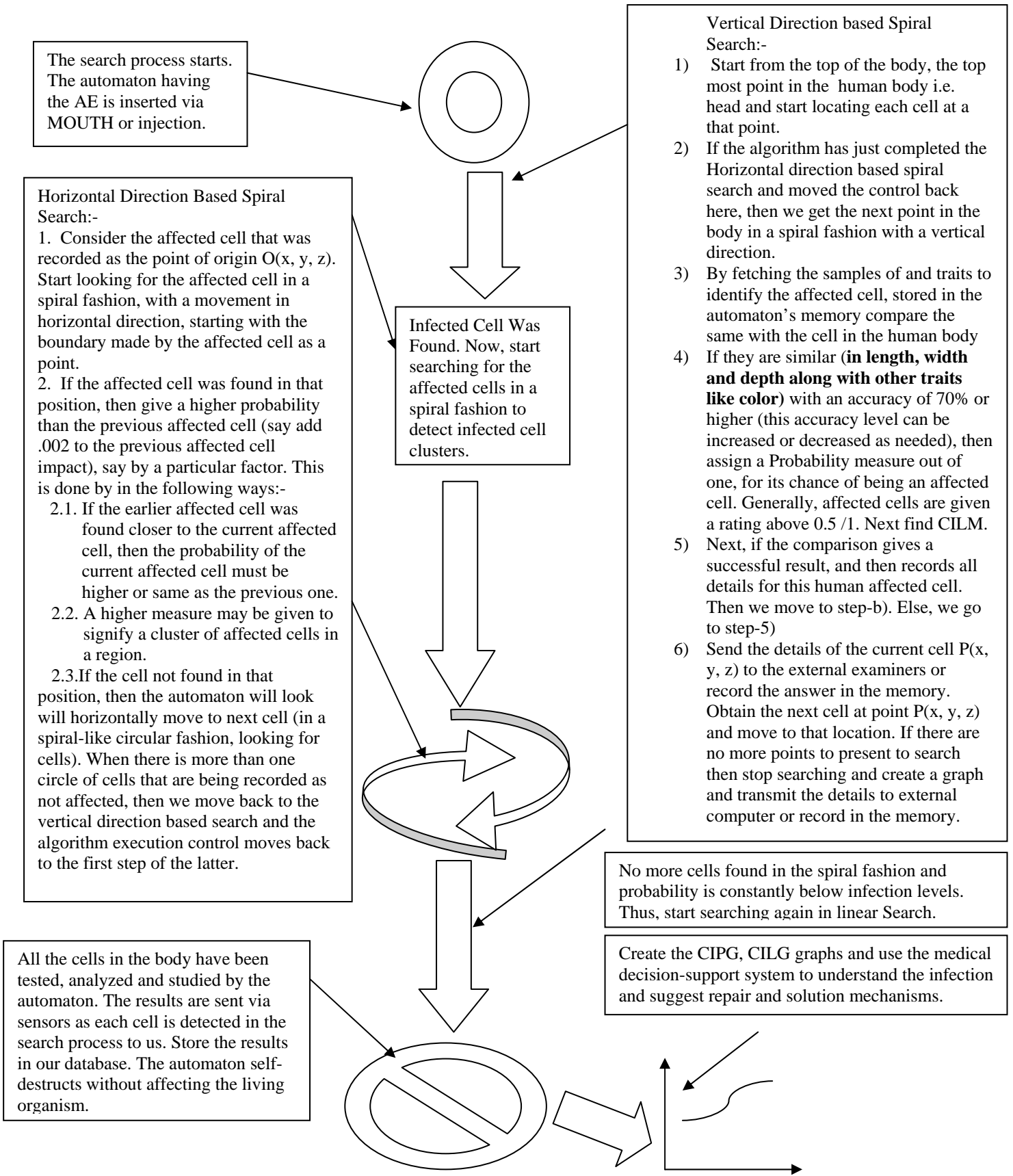

The search process starts. The automaton having the AE is inserted via MOUTH or injection.

Vertical Direction based Spiral Search:-
1) Start from the top of the body, the top most point in the human body i.e. head and start locating each cell at a that point.
2) If the algorithm has just completed the Horizontal direction based spiral search and moved the control back here, then we get the next point in the body in a spiral fashion with a vertical direction.
3) By fetching the samples of and traits to identify the affected cell, stored in the automaton's memory compare the same with the cell in the human body
4) If they are similar (**in length, width and depth along with other traits like color**) with an accuracy of 70% or higher (this accuracy level can be increased or decreased as needed), then assign a Probability measure out of one, for its chance of being an affected cell. Generally, affected cells are given a rating above 0.5 /1. Next find CILM.
5) Next, if the comparison gives a successful result, and then records all details for this human affected cell. Then we move to step-b). Else, we go to step-5)
6) Send the details of the current cell P(x, y, z) to the external examiners or record the answer in the memory. Obtain the next cell at point P(x, y, z) and move to that location. If there are no more points to present to search then stop searching and create a graph and transmit the details to external computer or record in the memory.

Horizontal Direction Based Spiral Search:-
1. Consider the affected cell that was recorded as the point of origin O(x, y, z). Start looking for the affected cell in a spiral fashion, with a movement in horizontal direction, starting with the boundary made by the affected cell as a point.
2. If the affected cell was found in that position, then give a higher probability than the previous affected cell (say add .002 to the previous affected cell impact), say by a particular factor. This is done by in the following ways:-
   2.1. If the earlier affected cell was found closer to the current affected cell, then the probability of the current affected cell must be higher or same as the previous one.
   2.2. A higher measure may be given to signify a cluster of affected cells in a region.
   2.3. If the cell not found in that position, then the automaton will look will horizontally move to next cell (in a spiral-like circular fashion, looking for cells). When there is more than one circle of cells that are being recorded as not affected, then we move back to the vertical direction based search and the algorithm execution control moves back to the first step of the latter.

Infected Cell Was Found. Now, start searching for the affected cells in a spiral fashion to detect infected cell clusters.

No more cells found in the spiral fashion and probability is constantly below infection levels. Thus, start searching again in linear Search.

Create the CIPG, CILG graphs and use the medical decision-support system to understand the infection and suggest repair and solution mechanisms.

All the cells in the body have been tested, analyzed and studied by the automaton. The results are sent via sensors as each cell is detected in the search process to us. Store the results in our database. The automaton self-destructs without affecting the living organism.

## Figure 4) The cellular Infection Level graph (CILG)

The average infection weight (CILM) is shown on y-axis and cell point P on x axis. This graph area shows all the cells and the degree of infection present in each of them. This also tells us if more than one type of infections are present in the body.

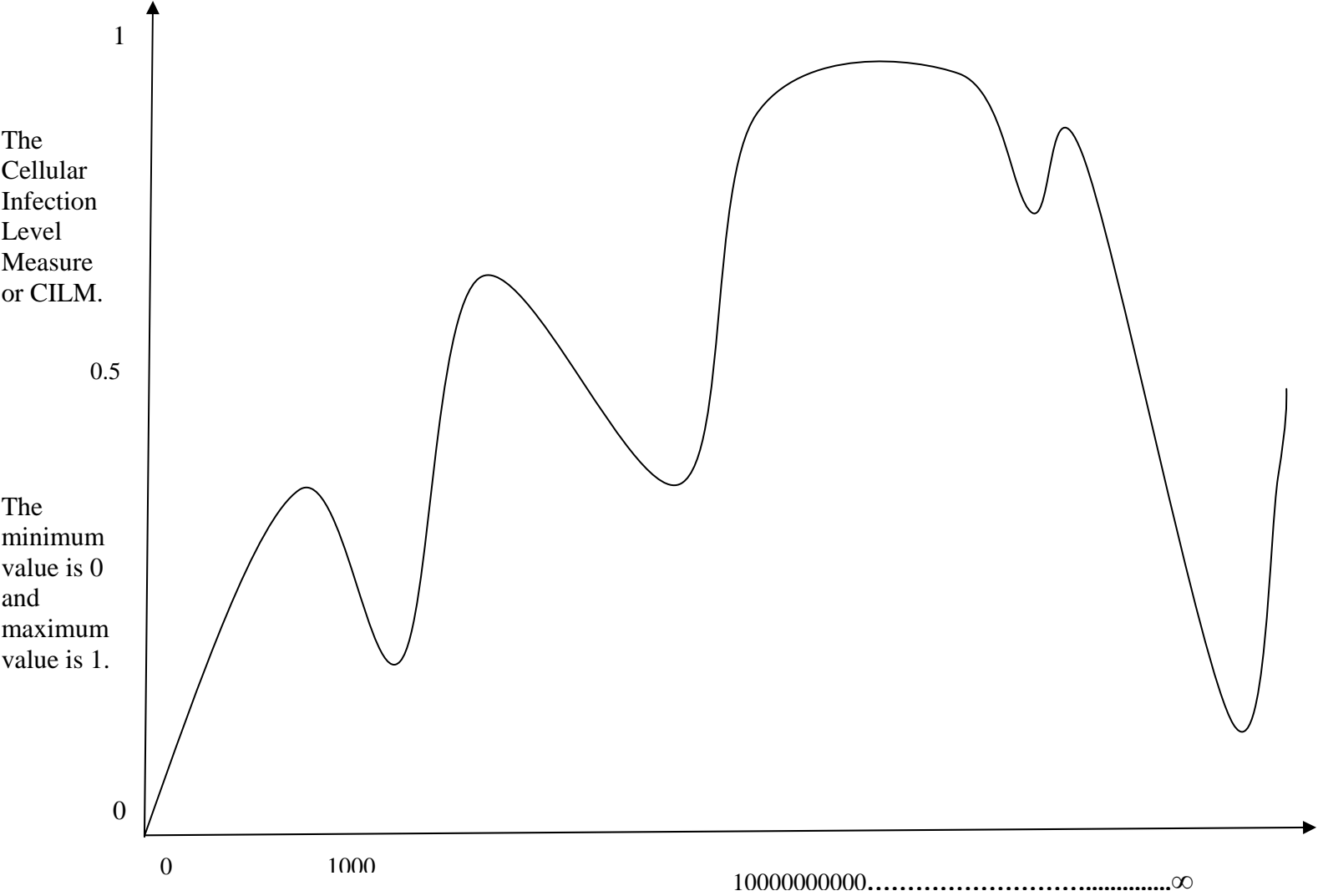

The Cellular Infection Level Measure or CILM.

The minimum value is 0 and maximum value is 1.

Each point P represents an individual cell in the body.

# Figure 5) CIPG graph for second type of infection.

Probability (CIPM) is shown on the y-axis and cell point P on the x axis. This graph area shows the cells in the human body that are affected or injured. For every type of infection found, there will be a new CIPM graph drawn. This graph is for the second type of infection found for the example in the paper. The graph will be plotted from the cell point where it was encountered. There will one CIPG graph for every new infection type. As shown in Figure-2, the corresponding probability for that point in all other CIPG graphs for other types of infection will be zero. Also, note that the point where the infection exists, a corresponding probability of zero is plotted on the first CIPM graph for the corresponding cell point. However, the CILM graph will remain the same for all the CIPM graphs.

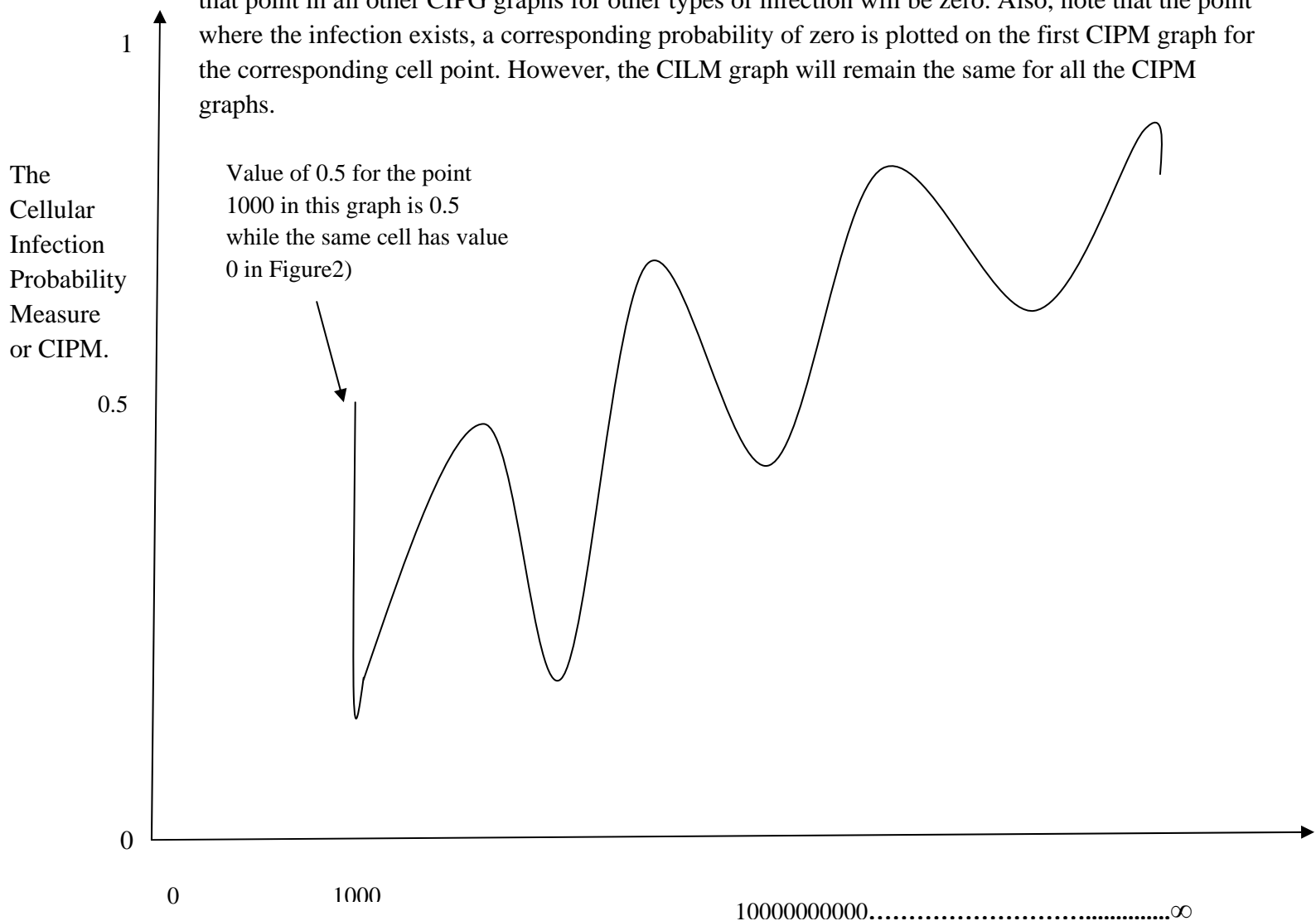

The Cellular Infection Probability Measure or CIPM.

Value of 0.5 for the point 1000 in this graph is 0.5 while the same cell has value 0 in Figure2)

Each point P represents an individual cell in the body.

# Figure 5) the cellular Infection Probability graph (CIPG) for second type of infection.

Probability (CIPM) is shown on the y-axis and cell point P on the x axis.This graph area shows the cells in the human body that are affected or injured. For every type of infection found, there will be a new CIPM graph drawn. This graph is for the second type of infection found for the example in the paper. The graph will be plotted from the cell point where it was encountered. There will one CIPG graph for every new infection type. As shown in Figure-2, the corresponding probability for that point in all other CIPG graphs for other types of infection will be zero. Also, note that the point where the infection exists, a corresponding probability of zero is plotted on the first CIPM graph for the corresponding cell point. However, the CIWM graph will remain the same for all the CIPM graphs.

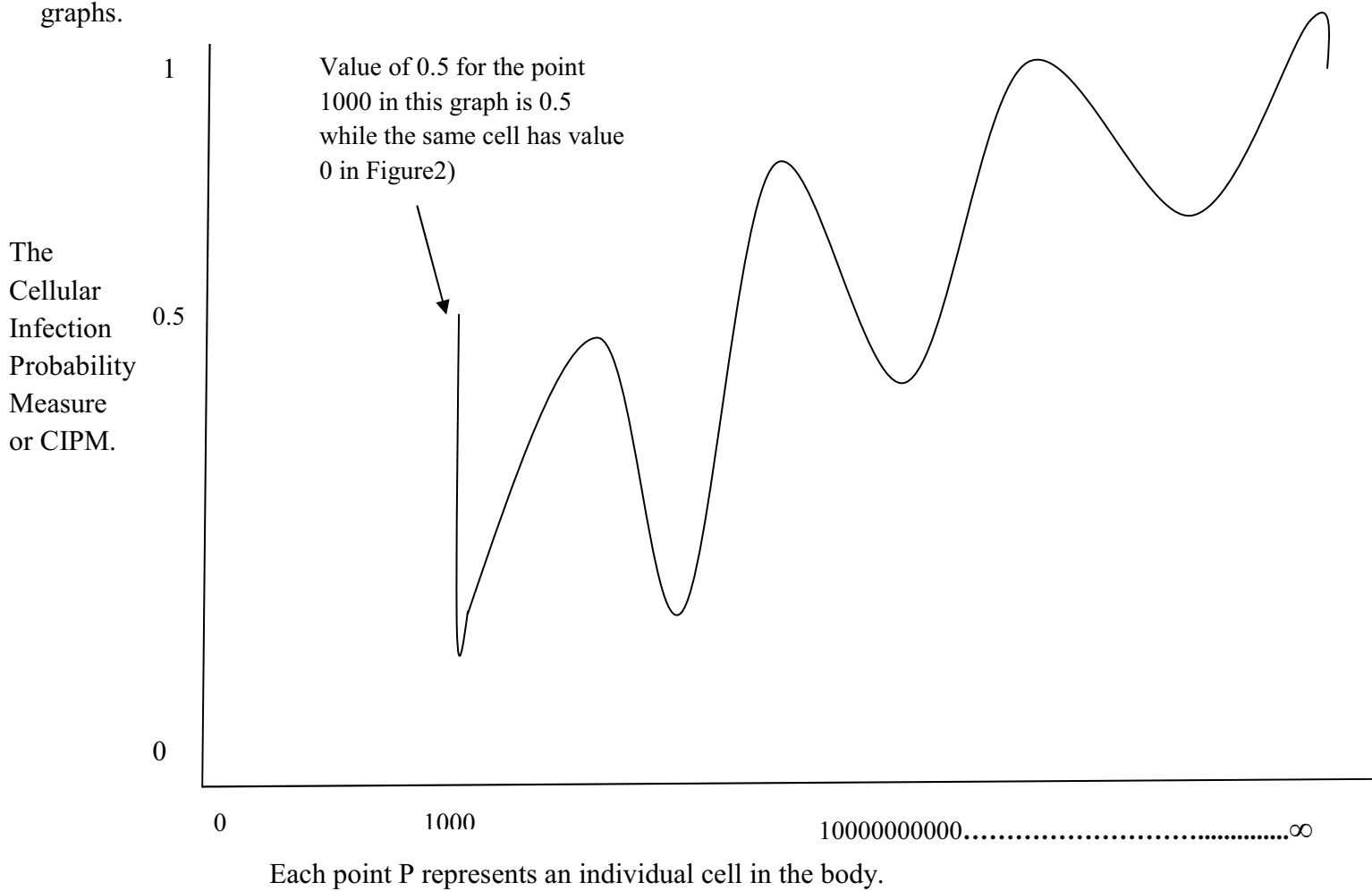

Each point P represents an individual cell in the body.

# Figure-2) The cellular impact probability graph (CIPG)

Probability (CIPM) is shown on the y-axis and cell point P on the x axis.

This graph area shows the cells in the human body that are affected or injured. Also, note that the point where the infection of another type exists, a corresponding probability of zero is plotted on this CIPM graph for the corresponding cell point. For e.g. in this CIPG graph the infected cell at 1000 is of probability zero as it is of different type of infection. The same cell point will have a probability of 0.5 in the other infection graph as shown in

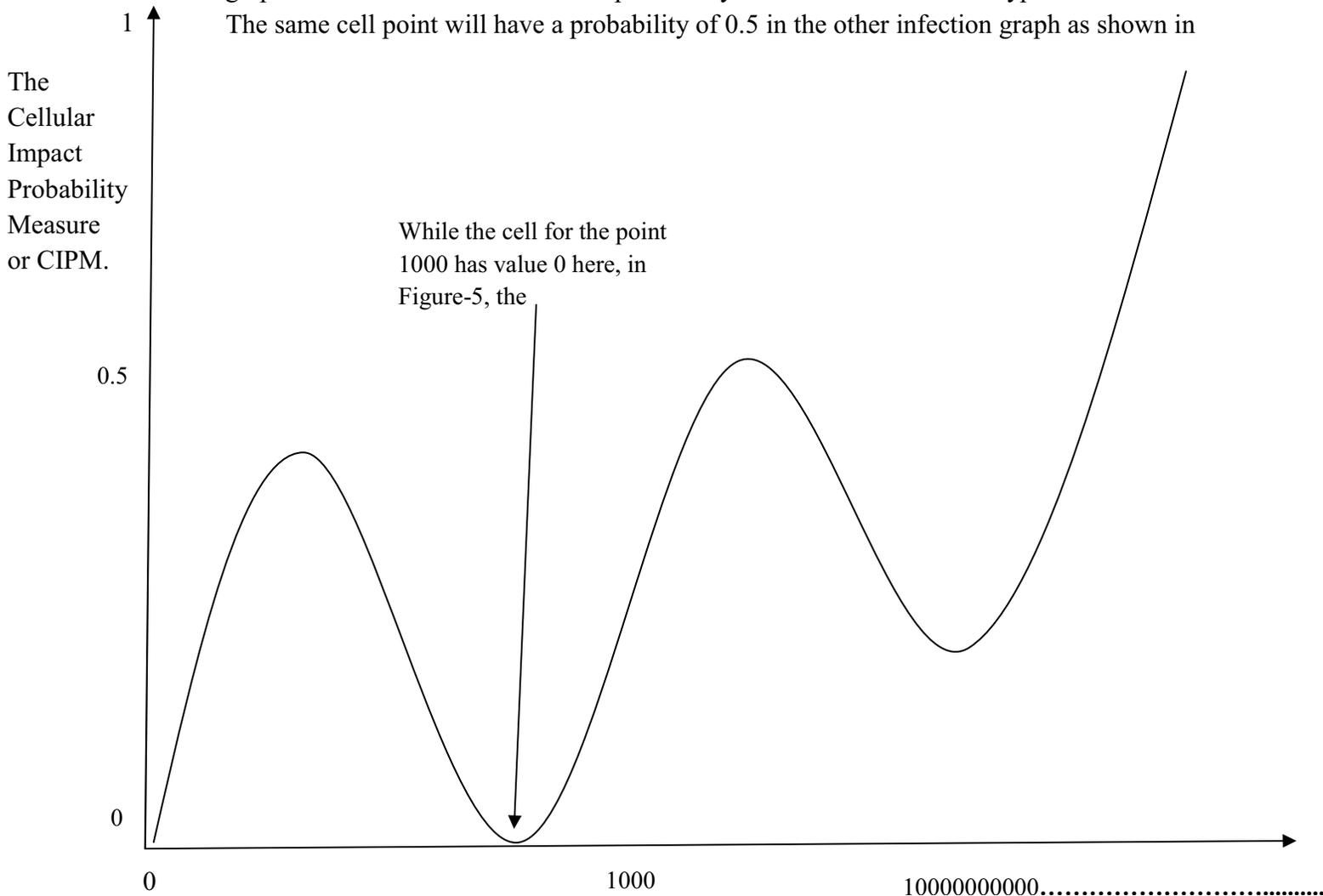

The Cellular Impact Probability Measure or CIPM.

While the cell for the point 1000 has value 0 here, in Figure-5, the

0.5

0

0          1000          10000000000..............................................

Each point P represents an individual cell in the body.

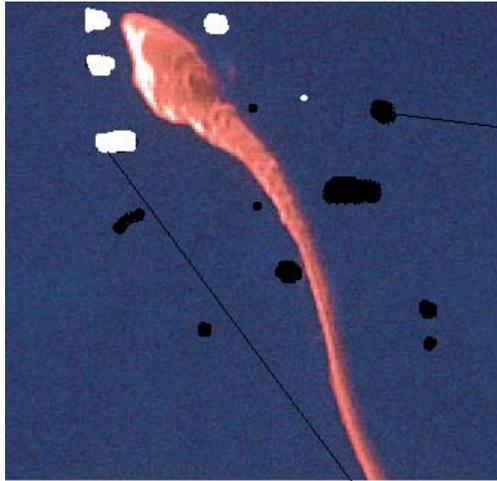

The spots in black anre the pathogens and external body agents that could attack and damage the automaton. PLease note that the automaton will first try to avoid the obstacles and if still not done, then it will try to attack the pathogen with its weapns in the head.

spots in white indicate the static objects that are barriers to the automaton.

The automaton may face internal objects too that are not enemies but may act as barriers in its progression. Again, these could also be the affected cells and thus they are recorded by the automaton before moving forward in its path.

**Figure-6-a) The human sperm as a possible automaton structure design, moving in a host body.**

The automaton will have a sperm-like
structure with the head containing the camera,
radar and processing unit.

The automaton will have a cover made up of biodegradable material so that it
can self-destruct easily without causing any problems to the host body.

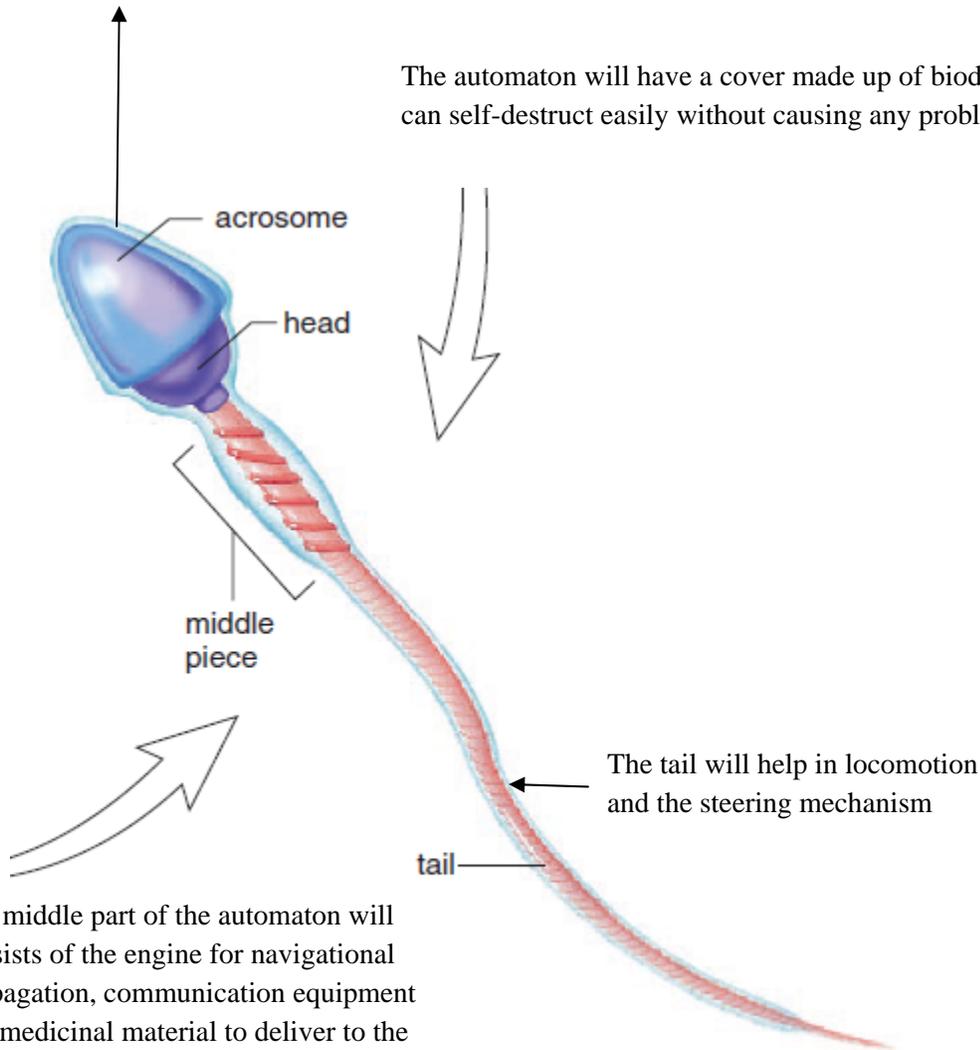

acrosome

head

middle
piece

The tail will help in locomotion
and the steering mechanism

tail

The middle part of the automaton will
consists of the engine for navigational
propagation, communication equipment
and medicinal material to deliver to the
affected cell.

**Figure- 6-b) The human sperm structure and its components.**

**Sources:- Mader. Human Biology. 7[th] Edition. The Mcgraw Hill Companies.
2001.**

**Figure-7) Experiment using NanoHive/HiveKeeper simulation software and simulated datasets.**

The experiments were undertaken as mentioned in the methodology section of this paper. The automatons used artificially intelligence based learning algorithm called IBSEAD for the help in detection using the framework. As and when obstacles were encountered, AIA model tried to avoid the obstacles (as programmed in this case. Here, the automatons can be programmed to destroy or put the medicine it carries at the obstacle. In this case, the automaton will put the medicine on the affected cells only post detection stage. In these scenarios, we have limited the experiments to the detection stage only)

1) **Initial Screen of the biological system using NanoHive and**

   **HiveKeeper: -** The point showed by x, y, z coordinates is the automaton. The tube like structure is being shown as the biological system in which as a certain section is visible here. The simulation was developed in the form of a nanotube like structure. The Initial Screen of the NanoHive simulation with 1 unknown entity is shown below:-

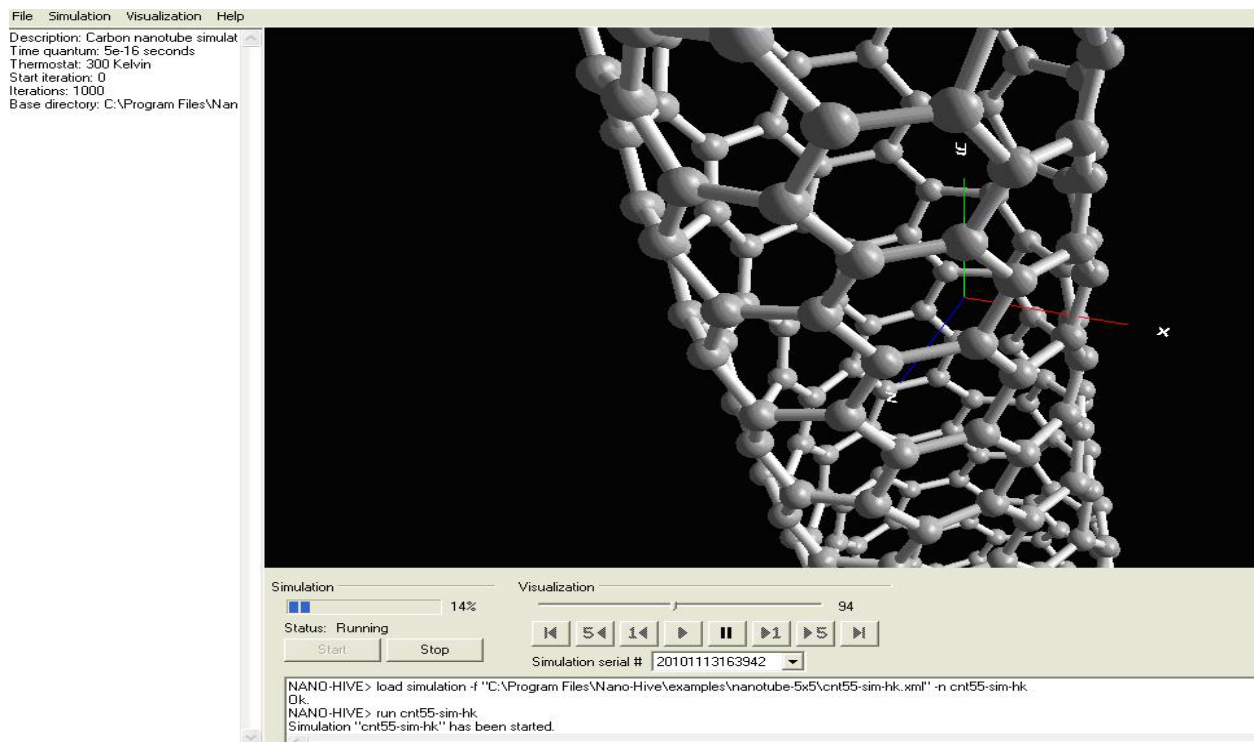

A similar process was performed for the rest of the test cases and by using the obstacle avoidance algorithms. One of the most interesting observations was the performance of the models in the presence of unknown entities. The use of IBSEAD (Dundas J and Chik D. 2010) helped in improving the performance of the AIA model, especially in complex scenarios.

2) In the next stages, we introduced certain regions as yellow which depicted affected cells. The red-coloured object is the obstacle which the automaton is avoiding and also collecting information about it as knowledge. The automaton tries to detect the affected cells using the framework.

**Yellow coloured regions are the affected regions in the biological system area. The biological system is the area in the cylindrical pipe.**

**The black coloured spot is the obstacle avoiding nano-object (Hla et al Nov 2008)**

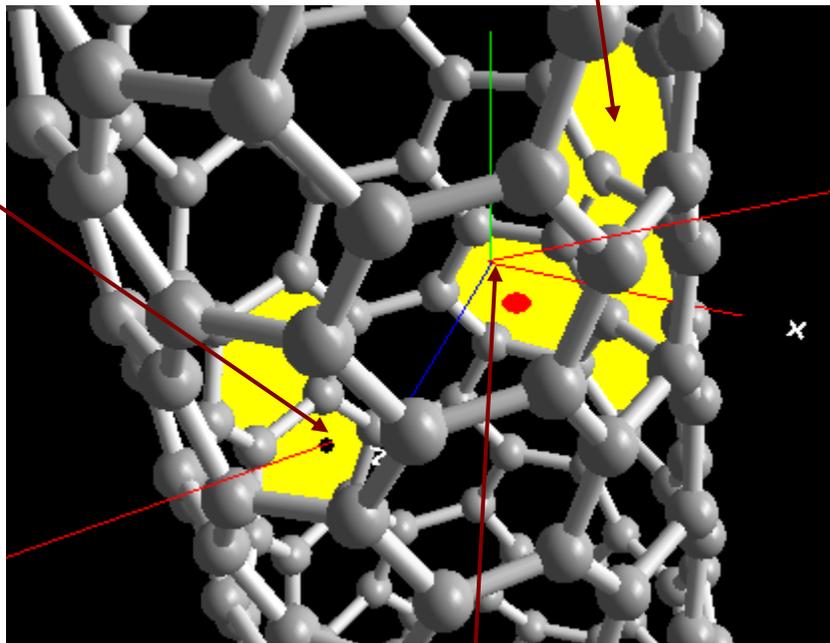

**Red coloured spot is the obstacle. Note how the black coloured nano-object tries to avoid the obstacle while the AIA based automaton tries to collect its information too.**

**The automaton is shown by x,y, and z coordinates. Note that while it is trying to avoid the obstacle using its hierarchical motion algorithm, it is also learning and collecting information about its obstacle (shown in red circular form between the yellow coloured affected regions.**

**Figure-8) Experiment Results using NanoHive/HiveKeeper simulation software and simulated datasets.**

Note that:-

1) Metric B: - Time Taken (which was faster - AIAM or OAM) for detection & avoiding the obstacle.

2) Metric C:- %age of the system's affected cells detected by OAM

3) Metric D :- %age of the system's affected cells detected by AIAM

4) As OAM is one of the best and common methods used till date, we have chosen OAM for our experiments here.

5) KE stands for known entities and UKE stands for unknown entities

Description: - The experiments were undertaken as mentioned in the methodology section of this paper.

Remarks: - Clearly from the results, the AIAM gives better performance against OAM. The results were quite noteworthy in the case of complex and multiple entities, especially for unknown entities.

| Experiment Cases | Were affected cells detected using | | Metric B (AIAM or OAM?) | Metric C | Metric D |
|---|---|---|---|---|---|
| | Obstacle Avoidance Model or OAM | AIA Model( using IBSEAD) | | | |
| 1 KE | Yes | Yes | OAM | **62** | **65** |
| 2 KEs | Yes | Yes | OAM | **67** | **69** |
| KEs > 10 | Yes (with lower success ratio) | Yes | AIAM | **58** | **71** |

| 1 UKE | Yes(with lower success ratio) | Yes | AIAM | **53** | **74** |
|---|---|---|---|---|---|
| 2 UKE | **No or Low** | Yes | AIAM | **23** | **76** |
| UKE > 10 | **No or Low** | Yes | AIAM | **33** | **76** |